\documentclass[11pt]{article}
\usepackage[a4paper,top=3cm,bottom=2cm,left=3cm,right=3cm,marginparwidth=1.75cm]{geometry}
\usepackage{amsmath,amssymb}
\usepackage{indentfirst}
\usepackage{lmodern}
\usepackage{amsfonts}
\usepackage{mathrsfs}
\usepackage{fancyhdr}
\usepackage{mathrsfs,epsfig,morefloats}
\usepackage{amsmath}
\usepackage{amsthm}
\usepackage{color}
\usepackage{lscape}
\usepackage{rotating}
\usepackage{caption2}
\usepackage{subfigure}
\usepackage{bm}
\usepackage{enumerate}
\usepackage{colortbl}
\usepackage[colorlinks,
linkcolor=blue,
anchorcolor=blue,
citecolor=blue]{hyperref}
\usepackage[table]{xcolor}
\usepackage{float}
\usepackage{array}

\begin{document}
\title{Model detection and variable selection for mode varying coefficient model}
\author{
~~ Xuejun Ma\thanks{ School of Mathematical Sciences, Soochow University, 1 Shizi Street, Suzhou, 215006, China. xuejunma@suda.edu.cn},
~~Yue Du\thanks{The corresponding author. School of Mathematical Sciences, Soochow University, 1 Shizi Street, Suzhou, 215006, China. Duuuuuyue@163.com}
~~Jingli Wang\thanks{School of Statistics and Data Sciences, Nankai University, Tianjin, China. jlwang@nankai.edu.cn}
}
\date{}
\maketitle

\begin{abstract}

Varying coefficient model is  often used in statistical modeling since it is  more flexible than the parametric model. However, model detection and variable selection of varying coefficient model are poorly understood in mode regression. Existing methods in the literature for these problems often based on mean regression and quantile regression.
In this paper, we propose a novel method to solve these problems for mode varying coefficient model based on the B-spline approximation and SCAD penalty. Moreover, we present a new algorithm to estimate the parameters of interest, and discuss the parameters selection for the tuning parameters and bandwidth. We also establish the asymptotic properties of estimated coefficients under some regular conditions. Finally, we illustrate the proposed method by some simulation studies and an empirical example.

\end{abstract}

\begin{quote}
\end{quote}

\begin{quote}
\noindent
{\sl MSC2010 subject classifications}: Primary 	62J07; secondary 	62G08.
\end{quote}

\section{Introduction}
\label{section1}
Suppose that $Y_i$ is a response variable and $(U_i,\boldsymbol{X}_i)$ is the associated covariate, $i=1,\ldots,n$, then the varying coefficient model (VCM) \cite{Hastie and Tibshirani1993} takes the form
\begin{align}
Y_i=\boldsymbol{X}_i^{\top} \boldsymbol{\alpha}(U_i) + \epsilon_i, \tag{1.1}
\end{align}
where $\boldsymbol{X}_i=(X_i^{(0)}, \dots, X_i^{(p)})^\top$ is the $(p+1)$ dimensional design vector with the first element $X_i^{(0)}=1$, and $\boldsymbol{\alpha}(U)=\{\alpha_0(U),\dots, \alpha_p(U)\}^{\top}$, $\alpha_j(U)$ is an unknown smooth function, $j=0,\ldots,p$. $U_i$ is an univariate index variable, without loss of generality, we assume $U$ ranges over the unit interval $[0,1]$, and $\epsilon_i$ is the random error.
VCM is attractive because the coefficients $\alpha_j(U)$'s depend on $U$, which not only reduces the modeling bias, but also avoids the ``dimensional disaster''. Under different assumptions, a large number of varying coefficient models are developed, such as additive model \cite{Hastie and Tibshirani1990}, partial linear model \cite{Hadle et al.2000}, single index function coefficient regression model \cite{Xia and Li1999}, etc.

In statistical modeling, variable selection is a significant issue. Penalized techniques have been proposed to conduct variable selection by shrinking coefficients of redundant variables to zero. There are various powerful penalization methods such as  LASSO \cite{Tibshirani1996}, SCAD \cite{Fan and Li2001}, adaptive LASSO \cite{Zou2006}, MCP \cite{Zhang2010}. Undoubtedly, variable selection is crucial and complex for the VCM. There are three types of variable selection problems for VCM:
 (1) detection of varying and constant coefficients \cite{Huang et al.2002}, (2) selection of variables which are nonzero with varying effects \cite{Cai et al.2000}, (3) selection of variables which are nonzero with constant effects \cite{Fan and Huang2005}. However, there are only a few methods on simultaneously solving all the three types of variable selection problems for VCM in the literature.  Tang, Wang, Zhu and Song \cite{Tang et al.2012} proposed a unified approach which is implemented by using two-step iterative procedure based on B-spline and adaptive Lasso for variable selection and the separation of varying and constant
 coefficients. Based on the adaptive LASSO penalty and Nadaraya-Watson estimator, Hu and Xia \cite{Hu and Xia2012} proposed a method for the semi-varying coefficient model, which can identify the constant coefficients and estimate the model simultaneously. Ma and Zhang \cite{Ma and Zhang2016} use local polynomial smoothing and the SCAD penalty to overcome the problems. Meanwhile, these approaches were built on mean regression or quantile regression, which are sensitive to non-normal errors and outliers. In this paper, we consider another important regression method--mode regression.

 Mode regression pays attention to the relationship between the majority data points and summaries the `most likely' conditional values, instead of mean or quantile \cite{Lee1989,Yao and Li2013}.  Yao, Lindsay and Li \cite{Yao et al.2012} studied the local modal regression for nonparametric models, which is robust when the data sets have heavy-tail or non-normal distributional error, and asymptotically efficient even if there are no outliers or the error follows normal distribution. Zhang, Zhao and Liu \cite{Zhang et al.2013}  discussed mode semiparametic partially linear varying coefficient model (MSPLVCM) based on the local polynomial method. However, they only showed the  selection of variables which are nonzero with constant effects. Zhao, Zhang, Liu, and Lv \cite{Zhao et al.2014} studied variable selection of MSPLVCM which used B-spline basis to approximate the varying coefficients, while this method requires users to detect the variables with varying or constant coefficients in advance by themselves.  To solve above problems,  based on  B-spline approximation and SCAD penalty, we propose a unified variable selection method, which is concerned with detecting varying and constant coefficients, selecting  variables with varying effects and constant effects for VCM using mode regression. This approach is motivated by the increasing statistical applications in biology, finance and neuroscience \cite{Zhu et al.2003, Li2010}.  Moreover, we also construct a two-step iterative procedure for the variable selection.

This paper is organized as follows. In section 2, we propose a new estimation method for VCM based on mode regression. We also introduce our method as well as algorithm in this section, which can solve the three types of variable selection problems in VCM simultaneously. Meanwhile, we also discuss the selection of bandwidth, interior knots, tuning parameters, and establish the asymptotic theories. Simulation studies and a real data analysis are presented in section  3. Finally, the article concludes in section 4. All proofs are contained in the Appendix.

\section{Main Results}
\label{section2}
We use the B-spline method to approximate unknown smooth function $\alpha_j(U), j=0,1,\ldots,p$. The B-spline basis functions with the degree of $d$ can be presented as $\tilde{\boldsymbol{B}}(U)=(B_1(U),\ldots, B_q(U))^{\top}$, where $q=k_n+d+1$ and $k_n$ is the number of interior knots. There exists a transformation matrix $\boldsymbol{G}$ \cite{Schumaker1981} such that $\boldsymbol{G}\tilde{\boldsymbol{B}}(U)=(1,\bar{\boldsymbol{B}}(U)^{\top})^{\top} \doteq \boldsymbol{B}(U)$. Then $\alpha_j(U)$ can be approximated by
\begin{align}
\alpha_j(U)\approx \boldsymbol{B}(U)^{\top} \boldsymbol{\gamma}_j = \gamma_{j1} + \bar{\boldsymbol{B}}^{\top}(U) \boldsymbol{\gamma}_{j*}, \qquad j=0,1,\ldots,p
\tag{2.1}
\end{align}
where $\boldsymbol{\gamma}_j=(\gamma_{j1},\boldsymbol{\gamma}_{j*}^{\top})^{\top}$ and %$\gamma_{j1}$ corresponds to the constant part of $\alpha_j(U)$,
$\boldsymbol{\gamma}_{j*}=(\gamma_{j2},\ldots,\gamma_{jq})^{\top}$.
%corresponds to the varying part of $\alpha_j(U)$.
Obviously, if $\vert\vert \boldsymbol{\gamma}_{j*} \vert\vert_{L_2}= (\sum_{k=2}^{q} \gamma_{jk}^2)^{1/2}=0$,  the variable $\boldsymbol{X}^{(j)}$ has only a constant effect.
Then based on mode regression,  we can estimate $\boldsymbol{\gamma}$ by maximizing %(\cite{Yao and Li2013})
\begin{align}\label{eq1}
  Q(\boldsymbol{\gamma}) = \sum_{i=1}^{n}K_h(Y_i - \boldsymbol{\Pi}_i^{\top} \boldsymbol{\gamma} )   \tag{2.2}
\end{align}
 where $\boldsymbol{\Pi}_i=(X_i^{(0)}\boldsymbol{B}(U_i)^{\top},\ldots, X_i^{(p)}\boldsymbol{B}(U_i)^{\top})^{\top}$,  $\boldsymbol{\gamma}=(\boldsymbol{\gamma}_0^{\top},\ldots,\boldsymbol{\gamma}_p^{\top})^{\top}$,  $K_h(\cdot)=\frac{1}{h}K(\cdot /h)$ and $K(\cdot)$ is a kernel function, and $h$ is the bandwidth. We assume $K(\cdot)$ is the standard normal density in this paper. Hence, the estimation of VCM can be transformed to a linear regression analysis.

\subsection{The penalized estimation via SCAD penalty}\label{subsec2.1}
In this section, we discuss the penalized mode VCM. First, the estimator $\boldsymbol{\gamma}$ can be obtained by
maximizing the following objective function
\begin{align}
Q(\boldsymbol{\gamma})-n\sum_{j=1}^{p}p_{\lambda_{1j}}(\vert \vert \boldsymbol{\gamma}_{j*} \vert \vert_{L_2}) - n \sum_{j=1}^{p} p_{\lambda_{2j}}(\vert \gamma_{j1} \vert) I(\vert \vert \boldsymbol{\gamma}_{j*} \vert \vert_{L_2}=0)   \tag{2.3}
\end{align}
It is equivalent to minimizing
\begin{align}\label{eq2}
l(\boldsymbol{\gamma})= - Q(\boldsymbol{\gamma}) + n\sum_{j=1}^{p}p_{\lambda_{1j}}(\vert \vert \boldsymbol{\gamma}_{j*} \vert \vert_{L_2}) + n \sum_{j=1}^{p} p_{\lambda_{2j}}(\vert \gamma_{j1} \vert) I(\vert \vert \boldsymbol{\gamma}_{j*} \vert \vert_{L_2}=0)  \tag{2.4}
\end{align}
where $\lambda_{1j}$  and $\lambda_{2j} >0$ are the tuning parameters, and
%given $a>2$,
 $p_{\lambda}(\cdot)$ is SCAD penalty function:
\begin{align}
p_{\lambda,a}(\theta)=\left\{
  \begin{array}{lcl}
  \lambda \theta & & {\theta \le \lambda}\\
  \frac{a \lambda \theta -0.5\left(\theta^2+\lambda^2\right)}{a-1}& & {\lambda < \theta \le a \lambda}\\
  \frac{\lambda^2\left(a^2-1\right)}{2\left(a-1 \right)} & & {\theta >a \lambda}
  \end{array}
  \right.
  \tag{2.5}
\end{align}
Fan and Li \cite{Fan and Li2001} suggested $a=3.7$.

 Now, we discuss two penalty terms in Eq.(\ref{eq2}). The frist one aims to identify varying and constant variables by penalizing the $L_2$ norm of the varying parts of each varying coefficient.
Tang et al \cite{Tang et al.2012} applied $L_{1} $ norm in the mean and quantile VCM. However,  $L_{1} $ norm does not treat $\boldsymbol{\gamma}_{j*} $ as a group, so it  is more inclined to set constant coefficients to varying coefficients.
The second term is designed to select zero constant effects. If  the variable $\boldsymbol{X}^{(j)}$ has a constant coefficient, all components of $\boldsymbol{\gamma}_{j*}$ are shrunk exactly to zero. On the other hand, if the variable $\boldsymbol{X}^{(j)}$ has no effect at all, both $\boldsymbol{\gamma}_{j*}$ and $\gamma_{j1}$ are shrunk to zero.
In practice, with the indicator function involved, it is very difficult to minimize the objective Eq.(\ref{eq2}), therefore, we implement a two-step iterative procedure to get the estimation of coefficients.

{\bf Step 1:}
Firstly, we obtain the $\hat{\boldsymbol{\gamma}}^{VC}$ by minimizing the penalized function
\begin{align}\label{eq3}
l_1(\boldsymbol{\gamma}^{VC})= - \sum_{i=1}^{n}Q(\boldsymbol{\gamma}^{VC}) + n\sum_{j=1}^{p}p_{\lambda_{1j}}(\vert \vert \boldsymbol{\gamma}_{j*}^{VC} \vert \vert_{L_2})
\tag{2.6}
\end{align}
%where $\lambda_{1j}$ are penalized parameters for $j$th varying coefficient function.
We penalize each varying coefficient by the $L_2$ norm of $\boldsymbol{\gamma_}{j*}^{VC}$.
 When the $j$th covariate has a constant effect, the $\boldsymbol{\gamma_}{j*}^{VC}$ is shrunk to zero, which is  an automatic separation of the varying and constant effects.

{\bf Step 2:}
After {\bf Step 1}, the model becomes a partially linear varying coefficient model.  Let $\boldsymbol{\gamma}_j^{CZ}=\boldsymbol{\gamma}_j^{VC}$ if  the variable $\boldsymbol{X}^{(j)}$ has a varying effect, $\boldsymbol{\gamma}_j^{CZ}=(\gamma_{j1}^{VC},\boldsymbol{0}_{q-1}^{\top})^{\top}$ if  the variable $\boldsymbol{X}^{(j)}$ has a constant effect. We obtain the estimator $\hat{\boldsymbol{\gamma}}^{CZ}$ by minimizing
\begin{align}\label{eq4}
l_2(\boldsymbol{\gamma}^{CZ})= - \sum_{i=1}^{n}Q(\boldsymbol{\gamma}^{CZ})+ n \sum_{j=1}^{p} p_{\lambda_{2j}}(\vert \gamma_{j1}^{CZ} \vert) I(\vert \vert \hat{\boldsymbol{\gamma}}_{j*}^{VC} \vert \vert_{L_2}=0)
\tag{2.7}
\end{align}
Here, we apply the same penalty $\lambda_{2j}$ to each constant coefficient part, and aim to exclude the irrelevant variables by assigning the SCAD penalty only to the terms that have been determined to be constant.

{\bf Step 3:}
Iterating {\bf Steps} 1 and 2 until convergence. Denote the final estimator of $\boldsymbol{\gamma}$ as $\hat{\boldsymbol{\gamma}}$, $\alpha_j(U)$ can be estimated by $\hat{\alpha}_j(U)=\boldsymbol{B}(U)^{\top} \hat{\boldsymbol{\gamma}}_j$ when  the variable $\boldsymbol{X}^{(j)}$ has a varying effect, and $\alpha_j(U)=\hat{\gamma}_{j1}$ when  the variable $\boldsymbol{X}^{(j)}$ only has a constant effect.

\subsection{Algorithms}\label{subsec2.2}
In this subsection, we extend the method proposed in \cite{Tang et al.2012} and the Modal Expectation-Maximization (MEM) method  \cite{Yao and Li2013} to minimizing Eq.(\ref{eq2}).

Because directly minimizing Eq.(\ref{eq2}) is difficult, we first locally approximate the penalty function $p_{\lambda}(\cdot)$ by a quadratic function proposed by Fan and Li \cite{Fan and Li2001}
\begin{align}
p_{\lambda}(\vert \omega \vert) \approx p_{\lambda}(\vert \omega_0 \vert) + \frac{1}{2}\frac{p_{\lambda}^{'}(\vert \omega_0 \vert)}{\vert \omega_0 \vert}(\omega^2-\omega_0^2),\qquad for\quad \omega \approx \omega_0
\tag{2.8}
\end{align}
Then, for a given initial value $\boldsymbol{\gamma}^{(0)}$, we can get
\begin{align}
p_{\lambda_{1j}}(\vert \vert \boldsymbol{\gamma}_{j*} \vert \vert_{L_2}) \approx  p_{\lambda_{1j}}(\vert \vert \boldsymbol{\gamma}_{j*}^{(0)} \vert \vert_{L_2}) +\frac{1}{2}\frac{p^{'}_{\lambda_{1j}}(\vert \vert \boldsymbol{\gamma}_{j*}^{(0)} \vert \vert_{L_2})}{\vert \vert \boldsymbol{\gamma}_{j*}^{(0)} \vert \vert_{L_2}}(\vert \vert \boldsymbol{\gamma}_{j*} \vert \vert_{L_2}^{2} - \vert \vert \boldsymbol{\gamma}_{j*}^{(0)} \vert \vert_{L_2}^{2})
\tag{2.9}
\end{align}

\begin{align}
p_{\lambda_{2j}}(\vert \gamma_{j1} \vert) \approx p_{\lambda_{2j}}(\vert \gamma_{j1}^{(0)} \vert) +\frac{1}{2} \frac{p^{'}_{\lambda_{2j}}(\vert \gamma_{j1}^{(0)} \vert)}{\vert \gamma_{j1}^{(0)} \vert}(\vert \gamma_{j1}\vert^{2} - \vert \gamma_{j1}^{0} \vert^{2})
\tag{2.10}
\end{align}
In this paper, $\boldsymbol{\gamma}^{(0)}$ can be chosen as the unpenalized estimator based on the least square regression
\begin{align}
\boldsymbol{\gamma}^{(0)}= \Big(\boldsymbol{\Pi}^{\top} \boldsymbol{\Pi}\Big)^{-1} \boldsymbol{\Pi}^{\top}\boldsymbol{Y}
\tag{2.11}
\end{align}
where $\boldsymbol{\Pi}=(\boldsymbol{\Pi}_1^{\top}, \ldots, \boldsymbol{\Pi}_n^{\top})^{\top}$. %Hence, with modified  MEM algorithm, %(\cite{Yao and Li2013}),
The algorithm mainly includes two steps, and each step has an EM iteration.
Therefore, we can obtain the sparse estimators  as follows:
\begin{enumerate}[(1)]
  \item {\bf Step 1:} Computation of  $\hat{\boldsymbol{\gamma}}^{VC(t+1)}$

  After the $t$-th iteration, we obtain $\hat{\boldsymbol{\gamma}}^{CZ(t)}$. Starting with $\hat{\boldsymbol{\gamma}}^{VC(t+1,0)} = \hat{\boldsymbol{\gamma}}^{CZ(t)}$,
  \begin{itemize}
  \item E-step:\\
  \begin{align}
  \pi(i|\hat{\boldsymbol{\gamma}}^{VC(t+1,m)})=\frac{K_h(Y_i- \boldsymbol{\Pi}_i^{\top}\hat{\boldsymbol{\gamma}}^{VC(t+1,m)})}{\sum_{i=1}^{n}K_h(Y_i- \boldsymbol{\Pi}_i^{\top} \hat{\boldsymbol{\gamma}}^{VC(t+1,m)} )}, \qquad i=1,\ldots,n
  \tag{2.12}
  \end{align}
  \item M-step:\\
  \begin{align}
  \begin{aligned}
  \hat{\boldsymbol{\gamma}}^{VC(t+1,m+1)} &= \arg \min_{\boldsymbol{\gamma}}\Big\{ \sum_{i=1}^{n}\{-\pi(i|\hat{\boldsymbol{\gamma}}^{VC(t+1,m)}) \log K_{h}(Y_i- \boldsymbol{\Pi}_i^{\top} \boldsymbol{\gamma})\} \\&+
  \frac{n}{2} \boldsymbol{\gamma}^{\top}  \boldsymbol{\Sigma}_{\lambda_1}(\hat{\boldsymbol{\gamma}}^{VC(t+1,m)})  \boldsymbol{\gamma}\Big\}\\
  &=\Big(\boldsymbol{\Pi}^{\top} \boldsymbol{W} \boldsymbol{\Pi} + n\boldsymbol{\Sigma}_{\lambda_1}(\hat{\boldsymbol{\gamma}}^{VC(t+1,m)})\Big)^{-1} \boldsymbol{\Pi}^{\top} \boldsymbol{W} \boldsymbol{Y}
  \end{aligned}
  \tag{2.13}
  \end{align}
  where
  \begin{align}
  \boldsymbol{\Sigma}_{\lambda_1}(\hat{\boldsymbol{\gamma}}^{VC(t+1,m)})={\rm diag}\Big\{ \boldsymbol{0}_q^{\top}, \Sigma_{\lambda_{11}}, \ldots, \Sigma_{\lambda_{1p}} \Big\}
  \tag{2.14}
  \end{align}
  with $\Sigma_{\lambda_{1j}} = \frac{p^{'}_{\lambda_{1j}}(\vert \vert \hat{\boldsymbol{\gamma}}_{j*}^{VC(t+1,m)} \vert \vert_{L_2})}{\vert \vert \hat{\boldsymbol{\gamma}}_{j*}^{VC(t+1,m)} \vert \vert_{L_2}} \{0, \boldsymbol{1}_{q-1}^{\top}\}^{\top},j=1,\ldots,p$, and $\boldsymbol{W}$ is an $n \times n$ diagonal matrix with diagonal elements $\pi(i|\hat{\boldsymbol{\gamma}}^{VC(t+1,m)})$.

  \item Iterating the E-step and M-step procedure until convergence reached and obtain $\hat{\boldsymbol{\gamma}}^{VC(t+1)}$, once $\vert \vert \hat{\boldsymbol{\gamma}}_{j*}^{VC(t+1)} \vert \vert_{L_2} < \epsilon$, we set $\hat{\boldsymbol{\gamma}}_{j*}^{VC(t+1)}= \boldsymbol{0}$. In the simulation, we set $\epsilon = 10^{-5}$, and the algorithm performs well.
  \end{itemize}

  \item {\bf Step 2:} Computation of $\hat{\boldsymbol{\gamma}}^{CZ(t+1)}$.

   After {\bf Step 1}, we set
  \begin{align}
  \hat{\boldsymbol{\gamma}}^{CZ(t+1,0)}=\left\{
  \begin{array}{rcl}
  & \hat{\boldsymbol{\gamma}}_{j}^{VC(t+1)}                           \quad &{\hat{\boldsymbol{\gamma}}_{j*}^{VC(t+1)} \ne \boldsymbol{0}}\\
  & (\hat{\gamma}_{j1}^{VC(t+1)},\boldsymbol{0}_{q-1}^{\top})^{\top}  \quad &{\hat{\boldsymbol{\gamma}}_{j*}^{VC(t+1)}  = \boldsymbol{0}}
  \end{array}
  \quad j=1,\ldots,p
  \right.
  \tag{2.15}
  \end{align}

  \begin{itemize}
  \item E-step:\\
  \begin{align}
  \pi(i|\hat{\boldsymbol{\gamma}}^{CZ(t+1,m)})=\frac{K_h(Y_i- \boldsymbol{\Pi}_i^{\top} \hat{\boldsymbol{\gamma}}^{CZ(t+1,m)})}{\sum_{i=1}^{n}K_h(Y_i- \boldsymbol{\Pi}_i^{\top} \hat{\boldsymbol{\gamma}}^{CZ(t+1,m)})}, \qquad i=1,\ldots,n
  \tag{2.16}
  \end{align}
  \item M-step:\\
  \begin{align}
  \begin{aligned}
  \hat{\boldsymbol{\gamma}}^{CZ(t+1,m+1)} &= \arg \min \Big\{\sum_{i=1}^{n}\{-\pi(i|\hat{\boldsymbol{\gamma}}^{CZ(t+1,m)}) \log K_{h}(Y_i- \boldsymbol{\Pi}_i^{\top} \boldsymbol{\gamma})\} \\&+
  \frac{n}{2} \boldsymbol{\gamma}^{\top}  \boldsymbol{\Sigma}_{\lambda_2}(\hat{\boldsymbol{\gamma}}^{CZ(t+1,m)})  \boldsymbol{\gamma}  \Big\}\\
  &=\Big(\boldsymbol{\Pi}^{\top} \boldsymbol{W} \boldsymbol{\Pi} + n\boldsymbol{\Sigma}_{\lambda_2}(\hat{\boldsymbol{\gamma}}^{CZ(t+1,m)})\Big)^{-1} \boldsymbol{\Pi}^{\top} \boldsymbol{W} \boldsymbol{Y}
  \end{aligned}
  \tag{2.17}
  \end{align}
  where
  \begin{align}
  \boldsymbol{\Sigma}_{\lambda_2}(\hat{\boldsymbol{\gamma}}^{CZ(t+1,m)})=diag\Big\{ \boldsymbol{0}_q^{\top}, \Sigma_{\lambda_{21}}, \ldots, \Sigma_{\lambda_{2p}}\Big\}
  \tag{2.18}
  \end{align}
  with $\Sigma_{\lambda_{2j}} = \frac{p^{'}_{\lambda_{2j}}(\vert \hat{\gamma}_{j1}^{CZ(t+1,m)} \vert)}{\vert \hat{\gamma}_{j1}^{CZ(t+1,m)} \vert} \{1, \boldsymbol{0}_{q-1}^{\top}\}^{\top},j=1,\ldots,p$, and $\boldsymbol{W}$ is an $n \times n$ diagonal matrix with diagonal elements $\pi(i|\hat{\boldsymbol{\gamma}}^{CZ(t+1,m)})$.

  \item Similar to {\bf Step1}, we iterate the EM procedure until convergence and obtain $\hat{\boldsymbol{\gamma}}^{CZ(t+1)}$, we also set $\hat{\gamma}_{j1}^{CZ(t+1)}= 0$ while $\vert \hat{\gamma}^{CZ(t+1)}_{j1}\vert < \epsilon$.
  \end{itemize}

  \item {\bf Step 3:} Iterating {\bf Steps} 1 and  2 until convergence.
\end{enumerate}

\subsection{Selection of parameters}\label{subsec2.3}
To implement the above method, we should determine the values of the tuning parameters $\lambda_{1j}$, $\lambda_{2j}$, the degree of B-spline $d$, the number of interior knots $k_n$, and the bandwidth $h$. In our numerical studies, we suggest using the cubic splines in which $d=3$.

We should choose values for $h$, $k_n$, $\lambda_{1j}$, $\lambda_{2j}$ in each iteration of the two-step procedure.  After the $t$-th iteration of the proposed two-step procedure, the details of selections are given as follows.
\begin{itemize}
\item Selection of bandwidth\\
We provide a bandwidth selection method for the practical use of the mode regression estimator. We first estimate $F(\boldsymbol{x},u,h) = E(K_h^{''}(\epsilon) | \boldsymbol{X}=\boldsymbol{x},U=u)
$, $G(\boldsymbol{x},u,h) = E((K_h^{'}(\epsilon))^2 | \boldsymbol{X}=\boldsymbol{x},U=u)$ by \cite{Yao et al.2012}
\begin{align}
\hat{F}(h) = \frac{1}{n}\sum_{i=1}^{n}K_h^{''}(\hat{\epsilon_i})
\tag{2.19}
\end{align}
\begin{align}
\hat{G}(h) = \frac{1}{n}\sum_{i=1}^{n}\{K_h^{'}(\hat{\epsilon_i})\}^2
\tag{2.20}
\end{align}
Then, we obtain $\hat{h}_{opt}$ by minimizing
\begin{align}
\hat{r}(h) = \frac{\hat{G}(h) \hat{F}(h)^{-2}}{\hat{\sigma}^2}
\tag{2.21}
\end{align}
where $
\hat{\epsilon}_i = Y_i- \boldsymbol{\Pi}_i^{\top} \hat{\boldsymbol{\gamma}}^{CZ(t)}$, $\hat{\sigma}^2=\frac{1}{n}\sum_{i=1}^{n}
\hat{\epsilon}_i^2$. The grid search method may be used to estimate $h$, according to the advise of Yao et al \cite{Yao et al.2012}, $h=0.5\hat{\sigma} \times 1.02^j,j=0,1,\ldots,l$ for some fixed $l$ such as $l=50$ or 100.

\item Selection of interior knots\\
We use SIC-type criterion to obtain $k_n$ by minimizing
\begin{align}
SIC_0(k)= - \log \sum_{i=1}^{n}Q(\hat{\boldsymbol{\gamma}}^{CZ(t)}) + \frac{\log n}{n}(v_m(k+d+1)+c_m)
\tag{2.22}
\end{align}
where $v_m$ denotes the number of covariates which have varying effects, $c_m$ denotes the number of covariates which have constant effects.

\item Selection of tuning parameter\\
We also use SIC-type criterion to obtain $\lambda_{1j}$, $\lambda_{2j}$ by minimizing
\begin{align}
SIC_1(\lambda_1)= - \log \sum_{i=1}^{n}Q(\hat{\boldsymbol{\gamma}}^{VC(t)}_{\lambda_1}) + \frac{\log n}{n}edf_1
\tag{2.23}
\end{align}

\begin{align}
SIC_2(\lambda_2)= - \log \sum_{i=1}^{n}Q(\hat{\boldsymbol{\gamma}}^{CZ(t)}_{\lambda_2}) + \frac{\log n}{n}edf_2
\tag{2.24}
\end{align}
where $\hat{\boldsymbol{\gamma}}^{VC(t)}_{\lambda_1}$ and $\hat{\boldsymbol{\gamma}}^{CZ(t)}_{\lambda_2}$ are the minimizers of Eq.(\ref{eq3}) and Eq.(\ref{eq4}) with
\begin{align}
\lambda_{1j}=\frac{\lambda_1}{\vert\vert \hat{\boldsymbol{\gamma}}^{VC(t)}_{\lambda_1 *} \vert\vert_{L_2}},j=1.\ldots, p
\tag{2.25}
\end{align}
and $edf_1,edf_2$ are defined as the total number of varying and nonzero constant coefficients \cite{Wang and Xia2009}.
\end{itemize}

\subsection{Asymptotic Properties}\label{subsec2.4}
Let $\{\alpha_j(\cdot),j=0,\ldots,v\}$ be the varying coefficients, $\{\alpha_j(\cdot),j=v+1,\ldots,s\}$ be constant coefficients and $\alpha_j(\cdot)=0,j=s+1,\ldots,p$. We assume some regular conditions.
\begin{enumerate}[(C1)]
  \item The varying coefficient functions $\alpha_j(\cdot),j=1,\ldots,v$ are $t$th continuously differentiable on $[0,1]$, where $t>2$.
  \item The density function $f_U(\cdot)$ is continuous and bounded away from zero and infinity on $[0,1]$.
  \item $\{\boldsymbol{X}_i, i=1,\ldots,n\}$ are uniformly bounded in probability, and the positive definite matrix $E(\boldsymbol{X}^{\top} \boldsymbol{X} |U=u)$ has bounded eigenvalues.
  \item The tuning parameters satisfy $n^{t/(2t+1)}\min\{\lambda_{1j},\lambda_{2j}\} \to \infty$, $\max\{\lambda_{1j},\lambda_{2j}\} \to 0$ as $n \to \infty$, and
\begin{align}
\lim \inf_{n \to \infty} \lim \inf_{\vert\vert \boldsymbol{\gamma}_{j*} \vert\vert_{L_2} \to 0^{+}} \frac{p_{\lambda_{1j}}^{'}(\vert\vert \boldsymbol{\gamma}_{j*} \vert\vert_{L_2})}{\lambda_{1j}}  >0 ,j=v+1, \ldots,p
  \tag{2.26}
\end{align}
\begin{align}
\lim \inf_{n \to \infty}\lim \inf_{\gamma_{j1} \to 0^{+}} \frac{p_{\lambda_{2j}}^{'}(|\gamma_{j1}|)}{\lambda_{2j}}  >0, j=s+1, \ldots,p
  \tag{2.27}
\end{align}

\end{enumerate}

{\bf Theorem 1.} Suppose conditions (C1)-(C4) hold and $k_n \sim n^{1/(2t+1)}$, with probability approaching 1, $\hat{\alpha}_j(\cdot)$ are nonzero constants, $j=v+1,\ldots,s$. And $\hat{\alpha}_j(\cdot)=0,j=s+1,\ldots,p $.

{\bf Theorem 2.} Suppose conditions (C1)-(C3) hold and $k_n \sim n^{1/(2t+1)}$, we have
\begin{align}
\vert\vert \hat{\alpha}_j(\cdot) - \alpha_j(\cdot) \vert\vert _{L_2}^2 =O_p(n^{-2t/(2t+1)} + a_n^2),j=0,\ldots,v
\tag{2.28}
\end{align}
where $a_n$ is defined as in Appendix.

\section{Numerical studies}
\subsection{Simulation}
We generate the random samples from following two models:

\textbf{Model 1}
\begin{align}
\begin{aligned}
Y_i &=15+20\sin(2\pi U_i) + (2-3\cos((6U_i-5)\pi/3))X_{i1} + (6-6U_i)X_{i2} +0.2X_{i3} +2X_{i4} +\delta_1 \epsilon_i
\end{aligned}
\tag{3.1}
\end{align}
where $U_i \sim Uniform(0,1)$. $X_{ij}$ comes from $N(0,1)$ for $j=1,\ldots,p$. $\delta_1=X_{i3}$.

\textbf{Model 2}
\begin{align}
\begin{aligned}
Y_i & =2\exp(1-U_i) + (1.5+3(\cos(2\pi U_i))^2)X_{i1} + (0.5+100U_i(1-U_i)(U_i-0.5))X_{i2}\\
& + (2-3\sin(2\pi U_i))X_{i3}+2X_{i4} +0.4X_{i5} -1.5X_{i6}+ \delta_2 \epsilon_i
\end{aligned}
\tag{3.2}
\end{align}
where $U_i \sim Uniform(0,1)$. $(X_{i1},\ldots,X_{ip})^{\top}$ are generated from a multivariate normal distribution with $Cov(X_{ik},X_{ij})=0.5^{\vert k-j\vert}$ for any $1 \le j,k \le p$, and $\delta_2=0.8X_{i5}$.

We considered the following four cases:
\begin{itemize}
\item Case 1: $\epsilon_i \sim N(0,1)$.
\item Case 2: $\epsilon_i \sim t(3) $.
\item Case 3: $\epsilon_i \sim Lap(0,1)$.
\item Case 4: $\epsilon_i \sim 0.5N(-1,2.5^2)+0.5N(1,0.5^2)$. $E(\epsilon)=0,mode(\epsilon)=1$.
\end{itemize}

The sample size $n$ is $500$. $p$ varies from 10 to 30. We repeat 500 times for each model, compare the performance of our proposed method (VCEM) with the basis expansion (BSE$_{L_1}$) based on mean regression \cite{Tang et al.2012}. Because the method using $L_1$ norm tends to set constant coefficients to varying coefficients,  we replace $L_2$ norm with $L_1$ norm for the BSE$_{L_1}$ and denote it as BSE$_{L_2}$. In order to evaluate the performance of the three methods, we consider the following criteria:
\begin{itemize}
\item SV: the average number of true varying coefficients (excluding the intercept) correctly select as varying coefficients.
\item SC: the average number of true constant coefficients correctly select as constant coefficients.
\item SZ: the average number of redundant variables correctly select as redundant variables.
\end{itemize}

\begin{table}[H]\centering
\caption{Simulation results for Model 1}
\label{table1}
%\resizebox{\textwidth}{!}{
\begin{tabular}{@{\extracolsep{5pt}} cccccc}
\\[-1.8ex]
\hline \\[-1.8ex]
$p$ & Case & Method & SV & SC & SZ  \\ \hline \\[-1.8ex]
10 &  \       &Oracle       &2.000  &2.000   &6.000\\
\  & 1        &BSE$_{L_1}$  &2.000	&1.616	&5.998\\
\  &       \  &BSE$_{L_2}$  &2.000	&1.672	&5.998\\
\  &       \  &VCEM         &2.000	&1.852	&5.998\\
\  & 2        &BSE$_{L_1}$  &2.000	&1.344	&5.970\\
\  &       \  &BSE$_{L_2}$  &2.000	&1.436	&5.946\\
\  &       \  &VCEM         &2.000	&1.668	&5.988\\
\  & 3        &BSE$_{L_1}$  &2.000	&1.398	&5.996\\
\  &       \  &BSE$_{L_2}$  &2.000	&1.558	&5.986\\
\  &       \  &VCEM         &2.000	&1.794	&5.992\\
\  & 4        &BSE$_{L_1}$  &2.000	 &1.274	 &5.972\\
\  &       \  &BSE$_{L_2}$  &2.000	 &1.284	 &5.966\\
\  &       \  &VCEM         &2.000	 &1.804	 &5.990\\
30 &  \       &Oracle       &2.000  &2.000      &26.000\\
\  & 1 &BSE$_{L_1}$  &2.000	&1.542	&25.936\\
\  &       \  &BSE$_{L_2}$  &2.000	&1.546	&25.906\\
\  &       \  &VCEM         &2.000	&1.746	&25.998\\
\  & 2  &BSE$_{L_1}$  &2.000	&1.304	&25.814\\
\  &       \  &BSE$_{L_2}$  &2.000	&1.408	&25.804\\
\  &       \  &VCEM         &2.000	&1.646	&25.824\\
\  & 3  &BSE$_{L_1}$  &2.000	&1.374	&25.948\\
\  &       \     &BSE$_{L_2}$  &2.000	&1.498	&25.804\\
\  &       \     &VCEM         &2.000	&1.706	&25.980\\
\  & 4   &BSE$_{L_1}$  &1.998	&1.268	&25.998\\
\  &       \                           &BSE$_{L_2}$  &1.996	&1.294	&26.000\\
\  &       \                           &VCEM         &1.998	&1.714	&25.996\\
\hline \\[-1.8ex]
\end{tabular}
%}
\end{table}

Tables \ref{table1} and \ref{table2} summarize the simulation results. From the simulation results, we can get the following comments:
\begin{enumerate}[(1)]
  \item In terms of identification of varying coefficients, VCEM is slight more efficient than BSE$_{L_1}$ and BSE$_{L_2}$. They perform similarly for selection of redundant variables.
  \item For the selection of constant coefficients, VCEM is superior to BSE$_{L_1}$ and BSE$_{L_2}$ because SC of VCEM is larger than the others.
  \item Even for the normal error case, VCEM  performs no worse than BSE$_{L_1}$ and BSE$_{L_2}$. Moreover, when the error distribution is asymmetric or has heavy tails, VCME is better than BSE$_{L_1}$ and BSE$_{L_2}$, since the mode regression will put more weight to the `most likely' data around the true value, which lead to robust and efficient estimator.
  \item BSE$_{L_2}$ is better than BSE$_{L_1}$, because $L_{2}$ norm takes the basis function as a group.

\end{enumerate}

\begin{table}[H]\centering
\caption{Simulation results for Model 2}
\label{table2}
%\resizebox{\textwidth}{!}{

\begin{tabular}{@{\extracolsep{5pt}} cccccc}
\\[-1.8ex]
\hline \\[-1.8ex]
p & Case & Method & SV & SC & SZ  \\ \hline \\[-1.8ex]
10 &  \       &Oracle       &3.000  &3.000      &4.000\\
\  & 1 &BSE$_{L_1}$  &2.986	&2.776	&3.998\\
\  &       \  &BSE$_{L_2}$  &3.000	    &2.906	&4.000\\
\  &       \  &VCEM         &3.000	    &2.994	&4.000\\
\  & 2   &BSE$_{L_1}$  &2.980	&2.762	&3.958\\
\  &       \  &BSE$_{L_2}$  &2.996	&2.722	&3.922\\
\  &       \  &VCEM         &3.000   	&2.960	&4.000\\
\  & 3  &BSE$_{L_1}$  &2.984	&2.738	&3.978\\
\  &       \     &BSE$_{L_2}$  &3.000	    &2706	&4.000\\
\  &       \     &VCEM         &3.000  	&2.974	&4.000\\
\  & 4    &BSE$_{L_1}$  &2.978	&2.660	&3.942\\
\  &       \                           &BSE$_{L_2}$  &2.988	&2.384	&4.000\\
\  &       \                           &VCEM         &2.994	&2.912	&3.998\\
30 &  \       &Oracle       &3.000  &3.000      &24.000\\
\  & 1  &BSE$_{L_1}$  &2.954	&2.594	&23.982\\
\  &       \  &BSE$_{L_2}$  &2.998	&2.946	&23.990\\
\  &       \  &VCEM         &3.000  &2.976	&24.000\\
\  & 2   &BSE$_{L_1}$  &2.962	&2.512	&23.942\\
\  &       \  &BSE$_{L_2}$  &2.982	&2.820	&23.879\\
\  &       \  &VCEM         &2.996	&2.948	&23.982\\
\  & 3  &BSE$_{L_1}$  &2.990	&2.716	&23.944\\
\  &       \     &BSE$_{L_2}$  &3.000	&2.702	&23.998\\
\  &       \     &VCEM         &3.000	&2.846	&24.000\\
\  & 4    &BSE$_{L_1}$  &2.968	&1.976	&24.000\\
\  &       \                           &BSE$_{L_2}$  &2.938	&2.324	&23.974\\
\  &       \                           &VCEM         &2.984	&2.884	&24.000\\
\hline \\[-1.8ex]
\end{tabular}
%}
\end{table}

\subsection{Application to the Boston housing data}

In this section, to illustrate the usefulness of the proposed procedure, we apply it to the Boston housing data. This data set concerns the median value of the owner-occupied homes, and it contains 506 observations on 14 variables, which can be found in the R package ``mlbench". We take MEDV (median value of owner-occupied homes in \$1000s) as the response variable. Covariates are as follows:
CRIM (per capita crime rate by town), ZN (proportion of residential land zoned for lots over 25,000 sq.ft), INDUS (proportion of non-retail business acres per town), CHAS (Charles River dummy variable (= 1 if tract bounds river; 0 otherwise)), NOX (nitrogen oxides concentration (parts per 10 million)), RM (average number of rooms per dwelling), AGE (proportion of owner-occupied units built prior to 1940), DIS (weighted mean of distances to five Boston employment centres), RAD (index of accessibility to radial highways), TAX (full-value property-tax rate per \$10,000), PTRATIO (pupil-teacher ratio by town), BLACK ($1000(Bk-0.63)^2$ where $Bk$ is the proportion of blacks by town), and LSTAT (lower status of the population (percent)) as the index variable. The response variable and covariates are standardized. The index variable LSTAT is transformed into interval $[0,1]$. We compare the results of variable selection and the mean of residual squares (MSE) for the three methods. The MSE can be calculated by
\begin{align}
MSE=\frac{1}{506}\sum_{i=1}^ {506}(y_i-\hat{y}_i)^2
\tag{3.3}
\end{align}

From Table \ref{table3}, the results of three methods are comparable. From Table \ref{table3}, we find AGE is an irrelevant variables; CRIM, INDUS and RAD have varying effects. While, PTRATIO has negative effects given by the BSE$_{L_2}$ and VCEM. RM has positive effects based on the BSE$_{L_1}$ and VCEM. DIS has negative effect based on the BSE$_{L_1}$ and BSE$_{L_2}$.  Moreover, the MSE of VCEM is smaller than the others.

\begin{table}[H]\centering
\caption{results for Boston housing data}
\label{table3}
\begin{tabular}{@{\extracolsep{5pt}} cccc} \\[-1.8ex] \hline \\[-1.8ex]
Variable & BSE$_{L_1}$ &BSE$_{L_2}$  & VCEM\\ \hline \\[-1.8ex]
CRIM    & V & V  & V  \\
ZN      & 0 & V  & V  \\
INDUS   & V & V  & V  \\
CHAS    & V & V & 0  \\
NOX     & V & 0 & -0.159 \\
RM      &  0.290&  V & 0.388 \\
AGE     & 0 &  0 & 0  \\
DIS     & -0.088 &  -0.023 & V  \\
RAD     & V & V  & V  \\
TAX     & 0 & 0  & V  \\
PTRATIO  & V & -0.038 & -0.192  \\
BLACK   & V & 0  & V  \\
\hline \\[-1.8ex]
MSE   & 0.202 & 0.202  & 0.198  \\
\hline \\[-1.8ex]
\end{tabular}
\end{table}

\section{Conclusion}
In this paper, we study the detection of varying coefficient models, selection of variables with nonzero varying effects and selection of variables with nonzero constant effects based on mode regression. We not only propose the VCEM algorithm but also establish asymptotic theories under some mild regular conditions. Obviously, numerical simulations show that our proposed method is effective. We would like to explore our proposed method to generalized varying coefficient models and so on for future study.

%\section*{Acknowledgments}
%The research was supported by the Natural Science Foundation of Jiangsu Province (Grants No. SBK2020040632).

\section*{Appendix}
Lemma A.1. Suppose conditions (C1)-(C4) hold, $\boldsymbol{\gamma}^{best}$ is the best approximation of $\boldsymbol{\gamma}$, and $\epsilon_1$, $a_1$, $a_2$ are positive constants, such that
\begin{enumerate}[(1)]
  \item $\vert\vert  \boldsymbol{\gamma}_{j*}^{best} \vert\vert_{L_2} >\epsilon_1,j=1,\ldots,v$, $\boldsymbol{\gamma}_j^{best}=(\gamma_{j1},\boldsymbol{0}_{q-1}^{\top})^{\top},j=v+1,\ldots,s$,
      $\boldsymbol{\gamma}_j^{best}=\boldsymbol{0},j=s+1,\ldots,p$.
  \item $\sup \vert \alpha_j(U) - \boldsymbol{B}(U)^{\top} \boldsymbol{\gamma}_j^{best} \vert  \le a_1 k_n^{-t}, j=1,\ldots,v$
  \item $\sup \vert \boldsymbol{\Pi}^{\top} \boldsymbol{\gamma}^{best} -\boldsymbol{X}\boldsymbol{\alpha}(U) \vert \le a_2k_n^{-t} $
\end{enumerate}

Lemma A.2. Let $\delta=O(n^{-t/(2t+1)})$. Define $\boldsymbol{\gamma}=\boldsymbol{\gamma}^{best}+\delta \boldsymbol{v}$, given $\rho >0$, there exists a large $C$ such that \cite{Zhao et al.2014}
$$P\{\sup_{\vert \vert \boldsymbol{v} \vert \vert=C} L(\boldsymbol{\gamma}) > L(\boldsymbol{\gamma}^{best})\} \le 1-\rho
$$

{\bf Proof of Theorem 2}

Similar to the proof of Theorem 1 in \cite{Zhao et al.2014}, let $a_n= \max_{j} \{\vert p^{'}_{\lambda_{1j}}(\vert\vert \boldsymbol{\gamma}_{j*} \vert\vert_{L_2}) \}$, $b_n= \max_{j} \{\vert p^{''}_{\lambda_{1j}}(\vert\vert \boldsymbol{\gamma}_{j*} \vert\vert_{L_2}) \}$. As $b_n \to 0$, we have
$\vert\vert \hat{\boldsymbol{\gamma}} -\boldsymbol{\gamma}^{best} \vert\vert = O_p(n^{-t/(2t+1)} + a_n)$. Therefore
\begin{equation*}
\begin{aligned}
&\vert\vert \hat{\alpha}_j(U) -\alpha_j(U) \vert\vert_{L_2}^2 \\
&= \sum_{i=1}^{n}(\hat{\alpha}_j(U_i) - \alpha_j(U_i))^2\\
&=\int_{0}^{1} (\hat{\alpha}_j(U) -\alpha_j(U))^2 dU \\
&=\int_{0}^{1} (\boldsymbol{B}(U)^{\top} \hat{\boldsymbol{\gamma}}_j - \boldsymbol{B}(U)^{\top} \boldsymbol{\gamma}^{best}_j +
\boldsymbol{B}(U)^{\top} \boldsymbol{\gamma}^{best}_j - \alpha_j(U))^2 dU \\
&\le 2 \int_{0}^{1} ((\boldsymbol{B}(U)^{\top} \hat{\boldsymbol{\gamma}}_j - \boldsymbol{B}(U)^{\top} \boldsymbol{\gamma}^{best}_j)^2 dU
+ 2 \int_{0}^{1} (\boldsymbol{B}(U)^{\top} \boldsymbol{\gamma}^{best}_j - \alpha_j(U))^2  dU\\
&= 2(\hat{\boldsymbol{\gamma}}_j - \boldsymbol{\gamma}^{best}_j)^{\top} \int_{0}^{1}{\boldsymbol{B}(U) \boldsymbol{B}(U)^{\top}}dU (\hat{\boldsymbol{\gamma}}_j - \boldsymbol{\gamma}^{best}_j)
+ 2 \int_{0}^{1} (\boldsymbol{B}(U)^{\top} \boldsymbol{\gamma}^{best}_j - \alpha_j(U))^2  dU
\end{aligned}
\end{equation*}
Because $\vert\vert \int_{0}^{1}{\boldsymbol{B}(U) \boldsymbol{B}(U)^{\top}} dU \vert\vert=O(1)$, we have
$$
(\hat{\boldsymbol{\gamma}}_j - \boldsymbol{\gamma}^{best}_j)^{\top} \int_{0}^{1}{\boldsymbol{B}(U) \boldsymbol{B}(U)^{\top}} dU (\hat{\boldsymbol{\gamma}}_j - \boldsymbol{\gamma}^{best}_j) =O_p(n^{-2t/(2t+1)} + a_n^2)
$$
And according to the Lemma A.1., we have
$$
\int_{0}^{1} (\boldsymbol{B}(U)^{\top} \boldsymbol{\gamma}^{best}_j - \alpha_j(U))^2  dU =O_p(n^{-2t/(2t+1)})
$$
Consequently, the proof has been completed.

$\hfill\qedsymbol$

{\bf Proof of Theorem 1}

By the property of SCAD, we know that $\max\{\lambda_{1j},\lambda_{2j}\} \to 0$ as $n \to \infty$, then $a_n=0$, then by Theorem 2, we have $\vert\vert \boldsymbol{\gamma} -\boldsymbol{\gamma}^{best} \vert\vert=O_p(n^{-t/2t+1})$

Firstly, if $\boldsymbol{\gamma}_{j*}=0$, it is clear that $\alpha_j(U)$ is a constant. If $\boldsymbol{\gamma}_{j*} \neq 0$, we have
\begin{equation*}\label{2}
\begin{aligned}
\frac{\partial l_1(\boldsymbol{\gamma})}{\partial{\boldsymbol{\gamma}_{j*}}} &=\sum_{i=1}^{n}K_{h}^{'}(Y_i-\boldsymbol{\Pi}_i^{\top} \boldsymbol{\gamma})X_i^{(j)}\bar{\boldsymbol{B}}(U) + \frac{np^{'}_{\lambda_{1j}}(\vert\vert \boldsymbol{\gamma}_{j*} \vert\vert_{L_2})}{\vert\vert \boldsymbol{\gamma}_{j*} \vert\vert_{L_2}} (sign(\gamma_{j,2})\gamma_{j,2},\ldots,sign(\gamma_{j,q})\gamma_{j,q})^{\top}\\
&=\sum_{i=1}^{n}\Big\{K_{h}^{'}(\epsilon_i + D_{ni})X_i^{(j)}\bar{\boldsymbol{B}}(U)
+ K_{h}^{''}(\epsilon_i+ D_{ni})X_i^{(j)}\bar{\boldsymbol{B}}(U)[\boldsymbol{\Pi}_i^{\top}( \boldsymbol{\gamma} - \boldsymbol{\gamma}^{best} )] \\
&+ K_{h}^{'''}(\eta_i)X_i^{(j)}\bar{\boldsymbol{B}}(U)[\boldsymbol{\Pi}_i^{\top}(\boldsymbol{\gamma} -\boldsymbol{\gamma}^{best})]^2 \Big\}
+ \frac{np^{'}_{\lambda_{1j}}(\vert\vert \boldsymbol{\gamma}_{j*} \vert\vert_{L_2})}{\vert\vert \boldsymbol{\gamma}_{j*} \vert\vert_{L_2}} (sign(\gamma_{j,2})\gamma_{j,2},\ldots,sign(\gamma_{j,q})\gamma_{j,q})^{\top}\\
&=n\lambda_{1j}\Big[O_p(\lambda_2^{-1} \bar{\boldsymbol{B}}(U)  n^{-\frac{t}{2t+1}}) +\lambda_{1j}^{-1}\frac{np^{'}_{\lambda_{1j}}(\vert\vert \boldsymbol{\gamma}_{j*} \vert\vert_{L_2})}{\vert\vert \boldsymbol{\gamma}_{j*} \vert\vert_{L_2}} (sign(\gamma_{j,2})\gamma_{j,2},\ldots,sign(\gamma_{j,q})\gamma_{j,q})^{\top}\Big]
\end{aligned}
\end{equation*}
where $\eta_i$ is between $Y_i-\boldsymbol{\Pi}_i^{\top} \boldsymbol{\gamma}$ and $\epsilon_i + D_{ni}$, $\epsilon_i=Y_i-\boldsymbol{X}_i^{\top} \boldsymbol{\alpha}(U_i)$, $D_{ni}= \boldsymbol{X}_i^{\top} \boldsymbol{\alpha}(U_i)- \boldsymbol{\Pi}_i^{\top}\boldsymbol{\gamma}^{best}$.

As we all know, $\sup_{u} \vert\vert \bar{\boldsymbol{B}}(U) \vert\vert=O(1)$, and by condition (C4) $n^{t/(2t+1)}\min\{\lambda_{1j},\lambda_{2j}\} \to \infty$ and $\lim \inf_{n \to \infty} \lim \inf_{\vert\vert \boldsymbol{\gamma}_{j*} \vert\vert_{L_2} \to 0^{+}} \frac{p_{\lambda_{1j}}^{'}(\vert\vert \boldsymbol{\gamma}_{j*} \vert\vert_{L_2})}{\lambda_{1j}}  >0 ,j=s+1, \ldots,p$, we prove that the sign of the derivation is completely determined by the second part of the derivation. Hence, with the Lemma A.2. $l_1(\boldsymbol{\gamma})$ gets its minimizer at $\hat{\boldsymbol{\gamma}}_{j*}^{VC} =0$, and $\hat{\alpha}_j(U)\approx \hat{\gamma}_{j1}^{VC} + \bar{\boldsymbol{B}}^{\top}(U) \hat{\boldsymbol{\gamma}}_{j*}^{VC} = \hat{\gamma}_{j1}^{VC} $, i.e. $\hat{\alpha}_j(U), j=v+1,\ldots,p$ are constants.

$\hfill\qedsymbol$

Secondly, since we have proved $\hat{\alpha}_j,j=v+1,\ldots,p$ are constant, we only need to prove $\hat{\gamma}_{j1}^{CZ}=0$ to obtain $\hat{\alpha}_j=0,$ for $j=s+1,\ldots,p$.
\begin{equation*}\label{2}
\begin{aligned}
\frac{\partial l_2(\boldsymbol{\gamma})}{\partial{\gamma_{j1}}} &=\sum_{i=1}^{n}K^{'}(Y_i-\boldsymbol{\Pi}_i^{\top} \boldsymbol{\gamma})X_i^{(j)}B_1(U) + np^{'}_{\lambda_{2j}}(\vert \gamma_{j1} \vert)sgn(\gamma_{j1})\\
&=\sum_{i=1}^{n}\Big\{K^{'}(\epsilon_i + D_{ni})X_i^{(j)}B_1(U) + K^{''}(\epsilon_i+ D_{ni})X_i^{(j)}B_1(U)[\boldsymbol{\Pi}_i^{\top}(\boldsymbol{\gamma} -\boldsymbol{\gamma}^{best})] \\
&+ K^{'''}(\eta_i)X_i^{(j)}B_1(U)[\boldsymbol{\Pi}_i^{\top}(\boldsymbol{\gamma} -\boldsymbol{\gamma}^{best})]^2 \Big\}
+ np^{'}_{\lambda_{2j}}(\vert \gamma_{j1}\vert)sgn(\gamma_{j1})\\
&=n\lambda_2[O_p(\lambda_{2j}^{-1} n^{-\frac{r}{2r+1}}) +\lambda_{2j}^{-1}p^{'}_{\lambda_{2j}}(\vert \gamma_{j1}\vert)sgn(\gamma_{j1}) ]
\end{aligned}
\end{equation*}
where $\eta_i$ is between $Y_i-\boldsymbol{\Pi}_i^{\top} \boldsymbol{\gamma}$ and $\epsilon_i + D_{ni}$, $\epsilon_i=Y_i-\boldsymbol{X}_i^{\top} \boldsymbol{\alpha}(U_i)$, $D_{ni}= \boldsymbol{X}_i^{\top} \boldsymbol{\alpha}(U_i)- \boldsymbol{\Pi}_i^{\top}\boldsymbol{\gamma}^{best}$.

By condition (C4) $n^{t/(2t+1)}\min\{\lambda_{1j},\lambda_{2j}\} \to \infty$ and $\lim \inf_{n \to \infty} \lim \inf_{\gamma_{j1} \to 0^{+}} \frac{p_{\lambda_{2j}}^{'}(|\alpha_{j1}|)}{\lambda_{2j}}  >0 ,j=s+1, \ldots,p$, we prove
$$\frac{\partial l_2(\boldsymbol{\gamma})}{\partial{\gamma_{j1}}} <0\quad when \quad  -\delta<\hat{\gamma}_{j1}^{CZ} <0$$
$$\frac{\partial l_2(\boldsymbol{\gamma})}{\partial{\gamma_{j1}}} >0\quad when \quad 0< \hat{\gamma}_{j1}^{CZ} < \delta$$
where $\delta=O(n^{-t/(2t+1)})$. Hence, $l_2(\boldsymbol{\gamma})$ gets its minimizer at $\hat{\gamma}_{j1}^{CZ} =0$, i.e. $\hat{\alpha}_j=0,j=s+1,\ldots,p$.

$\hfill\qedsymbol$

\end{document}